# Stock market's physical properties description based on Stokes' law

Geoffrey Ducournau [1]

**Abstract**

We propose in this paper to consider the stock market as a physical system assimilate to a fluid evolving in a macroscopic space subject to a Force that influences its movement over time where this last is arising from the collision between the supply and the demand of Financial agents. In fluid mechanics, this Force also results from the collisions of fluid's molecules led by its physical property such as density, viscosity, and surface tension.

The purpose of this article is to show that the dynamism of the stock market behavior can be explained qualitatively and quantitatively by considering the supply & demand collision as the result of Financial agents' physical properties defined by Stokes Law. The first objective of this article is to show theoretically that fluid mechanics equations can be used to describe stock market physical properties. The second objective based on the knowledge of stock market physical properties is to propose an econophysics analog of the stock market viscosity and Reynolds number ($N_R$) to measure stock market conditions, whether laminar, transitory, or turbulent. The Reynolds Number defined in this way can be applied in research into the study and classification of stock market's dynamics phases through for instance the creation of econophysics analog of Moddy diagram, this last could be seen as a physical way to quantify asset and stock index idiosyncratic risk. The last objective is to present evidence from a computer simulation that the stock market behavior can be a priori, and posteriori explained by physical properties (viscosity & density) quantifiable by fluid mechanics law (Stokes' law) and measurable with the stock market Reynolds Number.

**Keywords**: Econophysics, Fluid mechanics, Stokes Law, Reynolds Number, Computational model

## 1. Introduction

Fluid mechanics is a branch of continuous mechanics aiming to study fluid behaviors at a rest and in motion. Fluid mechanics had been applied to a wide range of disciplines such as oceanography focusing on ocean circulation [6], [7], [8], computational fluid dynamics focusing on numerical analysis [5], [12], [14], [16], geophysical & astrophysical [10], [11], meteorology focusing on the prediction of weather and meteorological phenomena [3]. More recently, because of the emerging uses of econophysics, analogies between the financial market and physical phenomena are more accepted by the academic community especially when dealing with social sciences that required theories describing collective phenomena, complex systems, and non-linearities. Therefore, analogies between fluid mechanics and the behavior of the financial market have been increasing within the econometric literature mainly to explain price fluctuations and price characteristics.

Wolfgang Breymann & Shoaleh Ghashghaie in their recent article [15] proposed to apply the Kolmogorov Cascade [2] to FX market to explain the heterogeneities between agents and their consequences on flow phenomenon at different time scales. They assumed an existing analogy between the turbulence in fluid mechanics arising from the transfer of energy of a fluid motion and the existing turbulences in the FX market from long to short temporal scales and emphasized this phenomena implication to the hierarchical structures in the FX market. They supposed that the pricing process can be explained as a hierarchical structure of collective phenomenon that shared significant features with hydrodynamic turbulence, in which there is also a self-similar hierarchy of vortices caused by a transfer of energy.

Moreover, A. Jakimowicz & J. Juzwiszyn [2] published a paper in 2014 in which they presented the stock market turbulences as the consequences of a physical balance behavior affected by the mutual interactions between the supply and the demand, creating consequently high-frequency dynamics of prices. The authors, by founding resemblance between the rotary trajectory observed in Hydrodynamics and the rotational dynamics of market movement in 3-dimensional space (time, price, volume) decided to consider in their paper the market vectors of volumes, prices, and time components as particles of a stream of liquid flowing through a pipe of a given cross-section. The purpose was to develop an economic equivalent of the Reynolds[3] number to explain the dynamics of stock exchange indices within this given 3-D space.

The idea developed in this paper first consists in considering the stock market dynamics as the consequences of a level of probability of having a collision between the supply and the demand, and secondly by trying to measure the stock market conditions (laminar, transitory, turbulent) through an economic analog of the Reynolds number. The innovative approach in our research relies on the fact that we have decided to not omit the intrinsic stock market physical properties at the origin of its intrinsic dynamism such as the viscosity and the density. Indeed, in their paper [2], A. Jakomiwicz and J. Juzwisyn argue that the equivalency of stock market viscosity has not been found in financial mathematics, therefore they propose a statistical approach to overcome this problem by using the J. Frenkel viscosity equation which is the ratio of a unit force to a diffusion process. In this paper, we decide to propose an economical analog of the stock market viscosity.

## 2. Stokes Law and fluid physical properties

George Gabriel Stokes was an Anglo-Irish physicist that spent most of his career in the study of fluid properties and mainly on the description of the motion of a sphere in a viscous fluid. His work led to the popularization of the "Stokes' Law", a theorem that describes mathematically the force required to move a sphere through a viscous fluid for a given velocity. This theorem is derived by linearizing the Navier-Stokes equation by considering advective inertial forces to be small compared to viscous forces. The Stokes' Law is written as,

$F_d = 6\pi\mu v R$ (1),

where $F_d$ is the drag force of the fluid, $\mu$ is the fluid viscosity, $v$ is the velocity of the sphere, and $R$ in this case is the radius of the sphere. A lot of interesting and useful derivation comes from equation (1) such as the calculation of the terminal velocity of a sphere moving to a fluid namely the velocity at the time the sum of forces applied to the sphere should be null. Stokes listed three forces acting on a sphere, $F_d$ the drag force, $F_b$ the buoyancy force, and $mg$ the gravity force where both forces $F_d$ and $F_b$ are acting in the same direction as opposed to the direction of the sphere gravitational acceleration. By summing forces, we can write,

$F_d + F_b - mg = 0$ (2),

at terminal velocity $v = v_T$ where $m = \rho_{fluid} V$ (3) and $F_b = \rho_{sphere} V g$ (4) with $\rho_{fluid}$ is the density of the fluid and $V = \frac{4}{3}\pi R^3$ the volume of the sphere, $\rho_{sphere}$ the density of the sphere and g the gravitational acceleration. By substituting (3) and (4) into (2) we get the following relationship,

$v_T = \frac{(\rho_{sphere} - \rho_{fluid})}{\mu} \frac{2}{9} R^2 g$ (2.1).

This relationship will be useful for part 3 of especially for giving an economic analog to the stock market viscosity. However, the equation (2.1) is valid in laminar flow namely when the fluid particles move along in smooth paths. However, in the case of a turbulent flow when the motion of fluid particles is random and irregular, the Stokes' Law becomes insufficient. In order to distinguish these two physical properties between laminar and turbulent flow, the scientist Osborne Reynolds [1] demonstrated the transition from laminar to turbulent flow can be quantified by the use of a dimensionless parameter today known as the Reynolds number ($N_R$) [9]. This number is a ratio between the inertial force and viscous forces within the fluid as follow,

$N_R = \frac{\rho \cdot v \cdot l}{\mu}$ (5)

where $\rho$ is the density of the fluid, $v$ the fluid velocity, $l$ the traveled length or in some cases can be the diameter of the system, and $\mu$ is the dynamic viscosity of the fluid. From observable experimentation, it is shown that laminar flow occurs when $N_R < 2300$ and turbulent flow occurs when $N_R > 2900$ and between both the fluid flow is considered as being in a transitory phase and unstable. In other words, the application of the Reynolds numbers to fluids problem is to determine the nature of the fluid flow conditions and for the case where we have viscous fluid, Stokes' Law is valid providing a Reynolds Number has a value less than 1.0.

---

[3] In 1883, the English scientist O. Reynolds demonstrated that a change of fluid flow could be determined with the use of the ratio of the inertial forces to the viscous forces. Inertial forces resist a change in the velocity of a physical system and lead to its movement whereas vicious forces are defined as resistance to flow. If vicious forces are dominant – the flow is laminar, if the inertial forces are dominant – the flow is turbulent.

Since turbulence displays both analogies with the time evolution of prices in the financial market and the evolution of fluid flow conditions [13], it seems interesting to consider the physical property of the stock market (viscosity and inertia) to reconstruct the econophysics analog of the Reynolds number and study the relationship between the different possible stock market physical properties and the evolution of its alternate flow conditions (laminar-turbulent).

### 3. Econophysics analog of stock market viscosity and Reynolds number

#### 3.1 Econophysics analog of the viscosity

The first objective of this paper is to propose an econophysics analog of the viscosity. To this end, we will consider the stock market as an entire system constituting of two elements (stock market obstacles and stock market fluid) both constituting of Financial agents with different properties. We will therefore start by giving the following definition:

- **Financial agent:**
  We define a Financial agent as a particle characterized by specific properties such as the size of the particle (number of shares or derivative contracts), the side of the particle (Buy-side or Sell-side), and the price of the particle. We consequently define every Financial agent as a particle of three coordinates [Side, Price, Size].

- **Stock market obstacles:**
  We define obstacles as preexisting Financial agents, that have taken a position on the stock market through an order to buy or sell at a specific price or better such as a "*Limit Order*" or a "*Stop Order*". They are considering pending order until the stock market price reaches their limit price, and they are registering within the stock market trading order-book. Consequently, every Financial agent that is registered within the order-book can be regarded as obstacle.

- **Stock market fluid:**
  We define a fluid as a particle corresponding for every time t to a Financial agent taking a position on the stock market. The fluid's properties are consequently changing and updating for every given time t accordantly to the evolution of the nature of financial agents taking a position. The fluid' properties can be defined by what we called in Finance a "*Market Order*" when the price of the particle corresponds to the current market price or a "*Limit Order/Stop Order*" when it corresponds to a specific price. As soon as the fluid interacts with obstacles, the Financial agents will be registered within the trading order-book and become an obstacle

The stock market flow is the result of interactions between the fluid and obstacle. Accordantly to the fluid's characteristics, we can distinguish two types of interactions:

- Active interaction: refer to a collision between the fluid and the order-book's Bid/Ask price. This interaction occurs when the fluid is defined as a "*Market Order*".
- Passive interaction: no collision between the fluid and order-book's Bid/Ask. This interaction occurs when the fluid price is defined as a "*Limit Order*" or a "*Stop Order*".

For a given time t, these interactions between the fluid and obstacles constitute the discrete properties of the stock market as the entire system.

We represent below an economic analog of a free body diagram of a stock market as being the interaction between the fluid and obstacles.

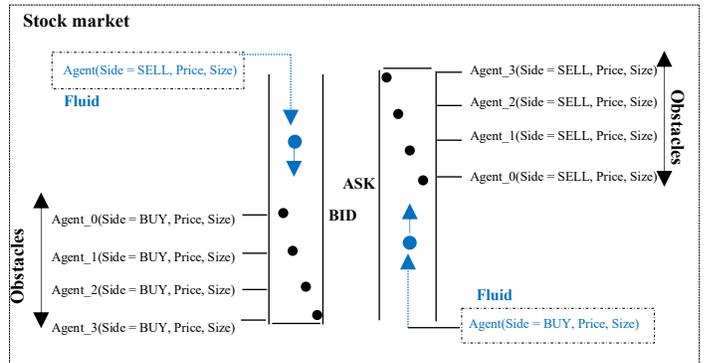

**Figure 1:** Free body diagram of a fluid interacting with obstacles.

Figure 1 can be seen as our entire system (stock market) for any given time t as being the interaction of a fluid (blue particle in Figure 1) and obstacles (preexisting agents with pending orders registering within the trading order-book). Later in this article, we will go deeply into this sketch describing our system because it turns out to be very useful when computing our simulation model.

By knowing the economic physical properties of the stock market and every new agent, we can also use the Stokes' Law to find an economic analog of the viscous parameter.

Indeed, with equation (2.1) and by omitting the radius and gravitational terms since agents (economics analogue of physical particles) remain electronic orders, we propose the following economic analog,

$$\mu_{fluid} = \frac{(\rho_{obstacle} - \rho_{fluid})}{v_T} \quad (2.2),$$

where $\mu_{fluid}$ is the dynamic viscosity of the fluid led by the new agent taking a position, $\rho_{obstacle}$ is the density of the preexisting agent with pending orders, $\rho_{fluid}$ is the density of the fluid function of the new agents 'properties and $v_T$ is the change in stock market price consequence of the interaction between the fluid and the obstacle.

We then define the density of the obstacle such that its value depends on the side order of the fluid. Indeed, if the new agent leading the fluid wants to "Buy" therefore the movement is led by a buying force and the resistance force is coming from the preexisting agents' properties present at the "ask" price level. The reciprocal is true. Therefore, we define the density such as:

$$\rho_{obstacle} = \frac{S_{ask} * P_{ask}}{V}$$

if the new fluid is led by a "buying" force and we define the stock market density such as:

$$\rho_{obstacle} = \frac{S_{bid} * P_{bid}}{V}$$

if the fluid is led by a "selling" force, where $S_{ask}$, $P_{ask}$, $S_{bid}$, $P_{bid}$, V correspond respectively to the size of the agents at the "ask" price level, the price value of the "ask", the size of the agents at the "bid" price level, the price value of the "bid" and the volume of transaction exchanged due to the new agent order. In the same way, we define the fluid density such as:

$$\rho_{fluid} = \frac{S_{order} * P_{order}}{V}$$

where $S_{order}$ corresponds to the size of the new agent order and $P_{order}$ corresponds to the price of its order. Consequently, by combining the previous relations to the equation (2.2) we obtain two possible fluid viscosity:
- If the fluid is led by a "buying" force:

$$\mu_{fluid} = \frac{S_{ask} * P_{ask} - S_{order} * P_{order}}{V * v_T} \quad (2.3)$$

- If the fluid is led by a "selling" force:

$$\mu_{fluid} = \frac{S_{bid} * P_{bid} - S_{order} * P_{order}}{V * v_T} \quad (2.4)$$

We understand from equations (2.3) and (2.4) that the viscosity of the fluid is a function of the collision between the supply and the demand. Indeed,
- if there is no collision then the viscosity tends to infinite because the terminal velocity will tend to 0 and it won't have any transaction exchanged, $V = 0$.
- if there is a perfect collision between the supply and the demand such that $S_{order} * P_{order} = S_{ask} * P_{ask}$ or $S_{order} * P_{order} = S_{bid} * P_{bid}$ then the viscosity is null, otherwise, the viscosity will depend on the strength of the collision and the change in stock market velocity due to the collision. To simplify the use of equation (2.3) and (2.4), we define the fluid viscosity as,

$$\mu_{fluid} = \frac{S_{onstacle} * P_{obstacle} - S_{order} * P_{order}}{V * v_T} \quad (2.5)$$

with $S_{obstacle} = S_{ask}$ and $P_{obstacle} = P_{ask}$ if the fluid is led by a "buying" force, or $S_{obstacle} = S_{bid}$ and $P_{obstacle} = P_{bid}$ if the fluid is led by a "selling" force.

### 3.2 Econophysics analog of the Reynolds number

Contrary to the viscosity, the literature review proposes some rewarding economic analog of the Reynolds number [2], [4], however, they skip some important physical property such as the existence of viscous and density parameters. Therefore, our second objective is to propose a new approach based on the derivation of the Reynolds number described in equation (5) such as,

$$N_R = \frac{\rho_{fluid} \cdot v_T \cdot l}{\mu_{fluid}} \quad (5.1)$$

We can assess here two trivial cases:
- The first one when there is no collision between the supply and the demand, in this case, the fluid viscosity tends to infinite and the Number of Reynolds tends to 0.
- The second one when we have a perfect collision as emphasized above, in this case, the fluid viscosity tends to zero and the Reynolds number tends to infinite.

Intuitively, we understand that the value of the Reynolds number will depend on whether we have a collision between the supply and the demand and the strength of each collision namely the change in price resulting from this collision. Therefore, we will define the conditional probability of having a perfect collision as,

$\mathbb{P} = P(S_{fluid} = S_{obstacle} | P_{fluid} = P_{obstacle}, \Omega_{side\_fluid})$ (6), with $\Omega_{side\_fluid}$ the sample space of the fluid side coordinate such that $\Omega_{side\_fluid} = \{buy, sell\}$.

$$\mathbb{P} = \frac{S_{fluid} * P_{fluid}}{S_{obstacle} * P_{obstacle}} \quad (6.1)$$

with $0 \leq \mathbb{P} \leq 1$. Consequently, by substituting (2.5) into (5) we obtain,

$$N_R = \frac{\frac{S_{fluid} * P_{fluid}}{V} \cdot v_T \cdot l}{\frac{S_{obstacle} * P_{obstacle} - S_{fluid} * P_{fluid}}{V * v_T}}$$

$$\Rightarrow N_R = \frac{S_{fluid} * P_{fluid} \cdot v_T^2 \cdot l}{S_{obstacle} * P_{obstacle} - S_{fluid} * P_{fluid}}$$

$$\Rightarrow N_R = \frac{\frac{S_{fluid} * P_{fluid}}{S_{obstacle} * P_{obstacle}} \cdot v_T^2 \cdot l}{1 - \frac{S_{fluid} * P_{fluid}}{S_{obstacle} * P_{obstacle}}} \quad (5.2)$$

And by combining (6.1) with (5.2) with obtain:

$$\Rightarrow N_R = v_T^2 * l * \frac{\mathbb{P}}{1 - \mathbb{P}} \quad (5.3)$$

with, $\frac{\mathbb{P}}{1-\mathbb{P}} = o_f$ the odds in favor of the probability of having a perfect collision such that:

$N_R = v_T^2 * l * o_f$ (5.4) with $\lim_{\mathbb{P}=1} o_f \to \infty$ and $\lim_{\mathbb{P}=0} o_f \to 0$.

By considering $l$ as the necessary travel length to make the stock market price change, we can define $l$ as being equal for any given time t to the value of the stock market spread. Consequently, we have defined an econophysics analog of the Reynolds number through the product of the square of the stock market flow with the stock market spread and a ratio that corresponds to a certain probability of having a collision between the fluid and the obstacle.

We assess that:
- If the probability $\mathbb{P}$ of having a perfect collision is equal to 0, then the Reynolds number tends to 0 and the stock market is not turbulent.
- If the probability $\mathbb{P}$ of having a collision is equal to 1, then the Reynolds number will tend to infinite and the stock market will be very turbulent or even chaotic.
- If the probability $\mathbb{P}$ is greater than 0 and smaller than 1, then the stock market turbulences will increase as $\mathbb{P}$ increases, and its speed to converge towards a laminar or turbulent flow's conditions will depend on the factor $v_T^2 * l$. Indeed, if the initial spread was large it will tend to create stock market jumps, large price changes, and excess volatility.

The innovative element through this economic analog of the Reynolds number is that the phenomenon of turbulence in the stock market can be viewed as the result of stock market physical properties and as the probability of having a collision between agents namely the probability that the supply meets the demand for any given time t.

4. **Computer simulation methodology and numerical results**

The construction of a numerical model is an important step to understand more precisely how different initial properties and mechanisms can lead to specific stock market dynamism and flow conditions (laminar, transitory, turbulent). As part of the study of a complex system, the development of the model is done by formalizing different behavioral rules that characterize the properties of the stock market such as the fluid which constitutes it (new Financial agent taking position within the trading order book) and that characterize the stock market behavior due to the interactions of the fluid with of others elements in his environment (obstacles, preexisting financial agents within the trading order book). The implementation of this model makes it possible to evaluate the dynamics of the system (stock market) in a collective context (interaction of heterogeneous & homogeneous Financial agents) namely evaluate the system viscosity and the Reynolds number.

### 4.1 Methodology

We recalled that we consider the stock market as a system constituting of two principal elements, a list of preexisting financial agents wishing to buy or sell for specific price registering within the trading order book (obstacles), and a fluid, a new financial agent taking a position and interacting with obstacles. After interacting, this new financial agent, in turn, becomes an element of the trading order book and leaves room for another Financial agent and so on.

Therefore, to create a computational model of the stock market as defined previously, we have to simulate for every time t, both the fluid and obstacles.

- Initialization of the stock market obstacles:

A trading order book refers to an electronic list of buy and sell orders organized by price level, commonly 20 price level (10 buy price level and 10 sell price level). Buy orders contain buyer's information including the bids price and the volume of contracts agents wish to purchase for every price level. We define this sample space of buyers' price level as $p_x \in \Omega_{price\_buyers}$ for $x \in \{-9, ..., 0\}$ with $p$ stands for price and with the element $\{p_0\}$ corresponding to the bid price.

Sell orders contain seller information including the ask price and the volume of contracts agents wish to sell for every price level. We define this sample space of sellers' price level as $p_x \in \Omega_{price\_sellers}$ for $x \in \{1, ..., 10\}$ with $p$ stands for price and with the element $\{p_1\}$ corresponding to the ask price.

The difference between the bid and the ask is called the spread. In our simulation, we randomly initialized the order book bid price level and the value of the spread. The purpose is being able to analyze the stock market dynamism according to different spread values. Moreover, by defining the bid price and the spread, we consequently obtain the list of price levels constituting the trading order book namely the sample space $\Omega_{price\_buyers}$ and $\Omega_{price\_sellers}$. Intuitively, we understand that the sample space of the buyers' and sellers' Price will evolve according to the interaction between the preexisting Financial agents within the order book and the stock market fluid.

Moreover, the volume of contract per price level namely the Size coordinate associated with every element that belong to the sample space $\Omega_{price\_buyers}$ and $\Omega_{price\_sellers}$ will be simulated according to a distribution function defined below.

- Initialization of the stock market fluid:

As we said, the fluid is constituting of a financial agent willing to interact with preexisting order book agents and is defined by three coordinates [Side, Price, Size]. In the model, the side of the Financial agent is defined by a given probability. For simplification, we decide to have an equiprobability of having a buy Side or sell Side order occurring on the market for any given time t such that the Side coordinate of the fluid has a sample space given by $\Omega_{side\_fluid} = \{buy, sell\}$ and the probability of any event that belongs to $\Omega_{side\_fluid}$ is $P(Buy) = P(Sell) = \frac{1}{2}$.

The Price coordinate of the Financial agent is rationally bounded by the existing list of price level constituting the order book for every given time t, such that the list of events corresponding to the fluid Price coordinate belongs to two possible sample spaces:

- if the Side coordinate of the fluid is the event {buy}:
$\Omega_{price\_fluid\_0} = \Omega_{price\_buyers} \cup \{p_1\}$

- if the Side coordinate of the fluid is the event {sell}:
$\Omega_{price\_fluid\_1} = \{p_0\} \cup \Omega_{price\_sellers}$

with every element that belongs to $\Omega_{price\_fluid\_0}$ and $\Omega_{price\_fluid\_1}$ is conditioning by a given probability. This probability corresponds to the likelihood of having a collision between the supply and the demand, and we defined previously this probability as
$\mathbb{P}_{collision} = P(P_{fluid} = P_{obstacle} \mid \Omega_{side\_fluid},)$, with:
$\mathbb{P}_{collision} = P(p_x = p_1 \mid \{buy\})$
$= P(p_x = p_0 \mid \{sell\})$.

Moreover, by defining L as the size of the sample space $\Omega_{price\_buyers}$ and $\Omega_{price\_sellers}$, the probability associated to each element of the respective sample space $\Omega_{price\_fluid\_0}$ and $\Omega_{price\_fluid\_1}$ are:

$P(p_x \mid \{buy\}) = \dfrac{1 - \mathbb{P}_{collision}}{L - 1}$ for $x \in \{-9; 0\}$

$P(p_1 \mid \{buy\}) = \mathbb{P}_{collision}$,

with $\sum_{x=0}^{9} P(p_x \mid \{buy\}) + P(p_1 \mid \{buy\}) = 1$
if the Side coordinate of the fluid is the event {buy} and,

$P(p_x \mid \{sell\}) = \dfrac{1 - \mathbb{P}_{collision}}{L - 1}$ for $x \in \{1; 10\}$

$P(p_0 \mid \{sell\}) = \mathbb{P}_{collision}$,
with $\sum_{x=1}^{10} P(p_x \mid \{sell\}) + P(p_0 \mid \{sell\}) = 1$

if the Side coordinate of the fluid is the event {sell}. We will consequently make simulations according to different probability $\mathbb{P}_{collision}$ of having a collision or not.

- Initialization of the Size coordinate of Financial agents

As we have explained previously, every financial agent whether the one corresponding to the stock market fluid, whether the ones corresponding to the order book are all defined by a coordinate vector [Side, Price, Size]. We have previously explained how the coordinates Price and Side are initialized within the model, remains now to explain how the coordinate Size will be defined. We decided to initialize the coordinate Size of every financial agent as a particle mass distributed in space according to a smoothing kernel with a smoothing length h such that we construct a Gaussian smoothing kernel with the function: $W(r; h) = \dfrac{1}{h^3 * \pi^{3/2}} e^{\left(\dfrac{-\|r\|^2}{h^2}\right)}$ (7), where $|r|$ is regarded as a distance between the Price coordinate of the fluid particle and the Prices coordinate constituting the trading order book bid and ask namely the element $\{p_0\} \in \Omega_{price\_buyers}$ and $\{p_1\} \in \Omega_{price\_sellers}$. Consequently, the Gaussian smoothing kernel enables us for a given Price coordinate of fluid to define its Size coordinate as the weighted average of neighboring observed preexisting Financial agents within the order book. In other words, the more the Price coordinate of a fluid Financial agent is far from the trading order book bid and ask, the more the value of its Size coordinate will be small. Moreover, the mass m of the Financial agent is defined as hyperparameter within the model which corresponds to the total amount of contracts exchanged during a trading session and the number of Financial agents that interact on the stock market for every given time t. This hyperparameter do not impact the dynamism of the stock market itself.

Given the mass distribution of every particle, we can reconstruct the Size coordinate at any location (here the location is referred to as the particle Price coordinate) using the smoothing kernels. For example, the Size coordinate at each Price coordinate is a sum over all the particles weighted by the distances between particles and the smoothing kernel: $Size_i = \sum_j mW(r_i - r_j; h)$ with $Size_i$ represents the Size coordinate of the fluid particle at time t, $r_i$ represents the Price coordinate of the fluid particle at time t, $r_j$ represents the neighbors observed particles within the trading order book ($\{p_0\}, \{p_1\}$) and $h$ is the smoothing kernel parameters also known as bandwidth and chosen as an hyperparameter within the model.

### 4.2 Computer simulation and numerical results

We propose through two different simulations to illustrate numerically the methodology emphasized previously. We initialize the hyperparameters with arbitrary values such that the bid Price coordinate $\{p_0\} = 3681$, the trading order book spread $|\{p_0\} - \{p_1\}| = 1$, the financial agent mass m = 2000, the smooth kernel parameter h = 10 and at every time t the probability of collision will be equal to $\mathbb{P}_{collision} = 0.99$ for the first simulation and to $\mathbb{P}_{collision} = 0.15$ for the second simulation.

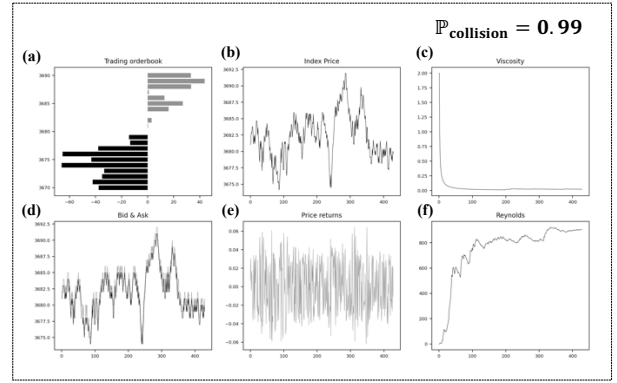

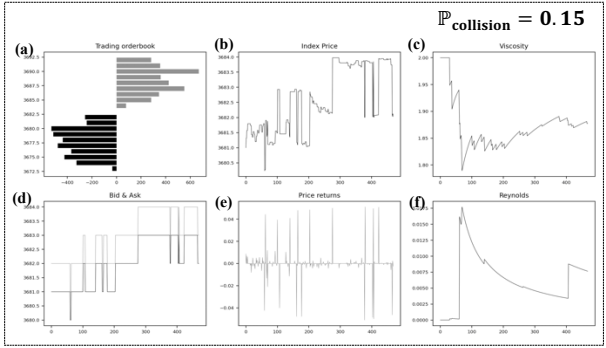

**Figure 2:** Computer simulation of stock market physical properties, (a) Trading order book, (b) Price motion, (c) Smoothing viscosity, (d) bid / ask motion, (e) Stock market returns, (f) Smoothing Reynolds Numbers.

Figure 2 illustrates the stock market properties outcoming by following the hyperparameter initialization, and after a period of 450 discrete time points.

It is important to mention that it is hardly possible to simulate and to represent graphically function that diverges to infinite numbers, which is the case for the viscosity. Therefore, for a question of practicality and readability, when the viscosity diverges to infinite, we have arbitrarily chosen to bound the viscosity value equal to 2. Consequently, if $\lim \mu \to \infty$ then we set $\mu = 2$. Moreover, in order to capture important patterns in the viscosity motion, we have decided to smooth the viscosity by divided for every time t its value by the maximum value among its series. In other words, whether we have $\mu = 2$ if the viscosity tends to infinite, whether we have $\mu = 1$ if the value corresponds to the maximum value of the series, whether $0 \leq \mu < 1$, if its value is smaller than the maximum value among the series. Then we have moving averaged the viscosity series. For the same purpose of capturing an important pattern in the Reynolds number, we have smoothing its series by moving averages.

Figure 2.(a) represents the trading order book after an iteration of 450 points, with the black horizontal histograms illustrating the Buyers price list level, the grey horizontal histograms illustrating the Sellers price list level, and the height of every histogram representing the cumulated volume of contracts.

Figure 2.(d) represents in black the evolution of the bid namely $\{p_0\}$ and in grey the ask namely $\{p_1\}$.

Figure 2.(b) and 2.(e) illustrate respectively the price motion and the price returns' motion. We can assess that for a very strong probability of collision such as $\mathbb{P}_{collision} = 0.99$, we obtain very rough stock market prices, with an excess volatility and conditional heteroskedasticity. We also have a very quick convergence of the viscosity towards 0 and a convergence of the Reynolds numbers towards 1000 meaning that regarding the conditions we have initialized, we have a stock market highly turbulent.

Reciprocally, when we have a weak probability of having a collision between the supply and the demand such as $\mathbb{P} = 0.15$, we assess a flattened price motion we weak volatility and weak amplitude. We also have a convergence of the viscosity towards the value of 2 namely towards infinite and convergence of the Reynolds numbers towards 0 meaning that regarding the conditions we have initialized, we have a stock market highly laminar.

We also assess that we have an inverse relationship between the viscosity and the Reynold number and that as much as the probability of having a collision between the supply and the demand increases, the Reynolds number tends to infinite and the viscosity to 0, and as much the probability decreases, the viscosity tend to infinite and the Reynolds number tends to 0.

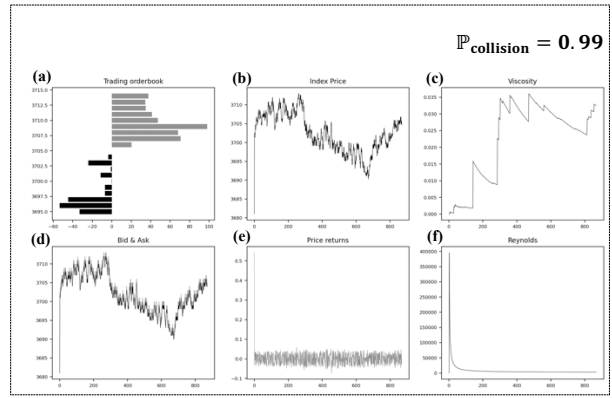

**Figure 3:** Computer simulation of stock market physical properties, (a) Trading order book, (b) Price motion, (c) Smoothing viscosity, (d) bid / ask motion, (e) Stock market returns, (f) Smoothing Reynolds Numbers.

Figure 3 illustrates the same stock market properties as Figure 2 with a probability of collision $\mathbb{P} = 0.99$ except that we have increased the initial spread such that $|\{p_0\} - \{p_1\}| = 20$. We can assess that the increase in the spread is a catalyzer of turbulences, roughness, excess volatility, and chaotic motions. Assessment that has been quantified through the increase of the Reynolds number to a very large value namely 4.10^5 and that finally converged to a value of roughly 6500.

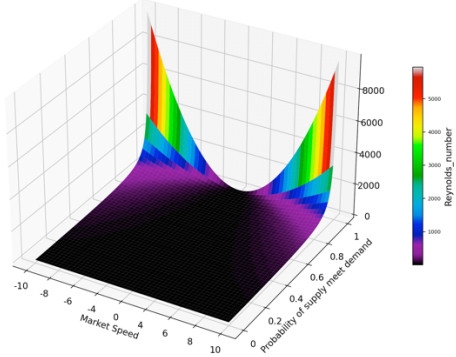

**Figure 4:** 3D Surface plot of Reynolds Numbers against market speed variation (change in prices for every time t) and the probability of having a collision $\mathbb{P}_{collision}$.

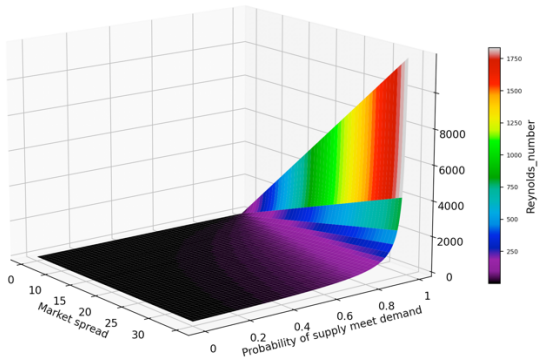

**Figure 5:** 3D Surface plot of Reynolds Numbers against market spread variation ( $l = |\{p_0\} - \{p_1\}|$ for every time t) and the probability of having a collision $\mathbb{P}_{collision}$.

Figure 4 illustrates the Reynolds number function of market speed and probability of collision. We can assess two local maxima when the probability of collision and when the absolute value of market speed is both maxima. Figure 5 illustrates the Reynolds number function of market spread and the probability of collision. We assess one local maximum, when the probability of having a collision is maximum and when the spread increases.

Consequently, through the computer simulations, we can observe the same property as the one given by equation (5.3), namely that the larger of the Reynolds number value is a function of the product of market speed, probability of having a collision between the supply and demand and is catalyzing by the larger of the spread value.

## 5. Conclusion

This article proposes a new approach to studying stock market dynamics through the computation of an econophysics analog of the Reynold number, a number that can be seen as a measure of stock market phases (laminar, transitory, turbulent).

First, we show evidence that the stock market can be seen as an entire system characterized by specific physical properties such as density and viscosity.

Second, we proposed a mathematical derivation of the stokes'Law taking into account the stock market physical properties. We finally show that the stock market Reynolds number for every given time t is a product of the square of market velocity to the market spread and to the odds in favor of the probability of having a perfect collision between the supply and the demand.

Third, we propose to develop a computational model that enables us to simulate the stock market as a complex system depending on a certain degree of freedom that represents the physical properties defined in the previous part. By manipulating these physical properties, the resulting simulations confirm the prediction given by the Reynolds number in part 3, namely that the level of stock market turbulences depend on the probability of having a collision between the supply and the demand and is catalyzed by the resulting market's price velocity and spread.

This paper can be applied in future econophysics research in particular as a way to propose a new approach of quantifying idiosyncratic risk on an asset or a stock index based on the measurement of its respective Reynolds Number.